\documentclass[12pt]{article}

\usepackage{amsmath,amssymb}
\usepackage{graphicx}
\usepackage{epstopdf}
\usepackage{epsfig}
\usepackage{array,multirow}

\begin{document}


\title{
\vspace*{-1cm}
\hfill{\normalsize\vbox{%
}}
\vspace*{3cm} \\
Gravitational Lorentz Violation \\
and Superluminality \\
via\\
 AdS/CFT Duality \vspace{0.5cm} \\
}

\author{
Raman Sundrum \\
 {\small \it Department of Physics and Astronomy, Johns Hopkins 
University}\\
{\small \it 3400 North Charles St., Baltimore, MD 21218 USA}
}

\date{}
\maketitle

\begin{abstract}
A weak quantum mechanical  coupling 
is constructed permitting superluminal communication within a 
preferred region of  a gravitating $AdS_5$ spacetime. 
This 
is achieved by adding a spatially non-local perturbation of a special kind to 
the Hamiltonian of a four-dimensional conformal field theory with 
a weakly-coupled $AdS_5$ dual, such as maximally 
supersymmetric Yang-Mills theory. In particular, two issues are given careful
treatment: (1) the UV-completeness of our deformed CFT, guaranteeing 
the existence of a ``deformed string theory'' AdS dual, and (2) the 
demonstration that superluminal effects can take place in AdS, both on its
boundary as well as in the bulk. Exotic Lorentz-violating properties such 
as these may have implications for tests of General Relativity, 
addressing the cosmological constant problem, or probing
``behind'' horizons. Our construction may give insight into 
the interpretation of wormhole solutions in Euclidean AdS gravity.
\end{abstract}

\newpage


\section{Introduction}

Relativistic invariance is a pillar
 of the fundamental laws of physics. It is worth questioning
 whether this structure is exact or just a (very good)
 approximation. The issue is subtle in the context of
 General Relativity which promotes Poincare invariance 
to a local symmetry, whose  breaking therefore requires  
some sort of Higgs mechanism. 
While low-energy effective field theories 
with partial Higgsing of General Relativity \cite{ted} \cite{ghost}, 
consistent with observation, 
have been constructed, their incorporation into UV-complete
 theories of quantum gravity such as string theory has not been 
demonstrated.

In the description in terms of a Higgs mechanism, relativistic invariance is 
respected by the 
dynamics and broken only by the state of some ``Higgs'' degrees of 
freedom. However, such a broad categorization encompasses some rather 
familiar and unremarkable cases.
 For example,
 the  preferred frame occupied by the cosmic microwave 
background can formally be thought of as spontaneously breaking Lorentz 
invariance, and by going to co-moving coordinates 
general coordinate invariance
is effectively ``Higgsed''.   But there may also be exotic Higgs phases,  
 breaking relativistic invariance with much more dramatic
implications.
 There is of course the possible phenomenology of
measurable quantitative deviations from standard expectations of General 
Relativity. See Refs. \cite{pheno} for examples.
But there may be important qualitative effects as well. In Lorentz 
invariant theories, superluminal 
propagation and interactions in one reference frame would imply 
acausal effects in other frames. But this need not be the case for 
Lorentz-violating interactions, which may have a preferred frame in which 
causal unitary evolution is defined. 
 Superluminal interactions would be liberating in our vast universe, and
might also allow us to probe ``behind'' black hole  and cosmological horizons, 
normally off limits by relativistic causality. If Lorentz violation is 
significant, it can go beyond being merely a probe of horizons, it can modify 
their character \cite{bh}. The   
observation that some apparently ``innocent'' effective field theories 
display superluminal behavior \cite{nicolis} would no longer immediately be a
disqualification. Lorentz-violation in 
General Relativity  may also help us understand some of 
gravity's other mysteries. For example, in Ref. \cite{us} it was shown that 
large Standard Model quantum contributions to dark energy can be canceled 
by a symmetry, ``Energy-Parity'', but the
 longevity of flat empty space then 
requires a Lorentz-violating short-distance breakdown of General Relativity.
Finally, if relativistic invariance
 is an approximation, it may well be an emergent 
(accidental) symmetry, simple examples of which occur in the long-wavelength 
approximation of some condensed matter systems.
 The question then arises, what more fundamental 
structure or symmetry underlies Relativity.

In this paper, we exploit the powerful approach to quantum gravity offered by 
the AdS/CFT correspondence \cite{adscft} (reviewed in Ref. \cite{adscftrev}), 
to engineer UV-complete gravitational dynamics 
exhibiting weak breaking of (local) Poincare invariance and superluminal 
action-at-a-distance. The construction is made on the CFT side of the 
correspondence, specifically by perturbing strongly-coupled 
large-$N_{color}$ ${\cal N} = 4$ 
supersymmetric Yang-Mills (SYM) theory by suitably chosen spatially 
non-local operators.
The advantage of working in terms of these holographic degrees of freedom is 
that it finesses
 the tricky issues of breaking  gauge symmetries, such as 
general coordinate invariance, that appear in the dual description of 
AdS gravity. In particular, the correspondence relates the breaking of 
gauge symmetry in AdS
to breaking of global symmetry in the CFT, which is much easier to understand. 
This feature is illustrated by two well-known examples (in which, however, 
the relevant Higgs dynamics in AdS takes quite 
familiar and unexotic forms). 
The first example is given by simply
 adding a spacetime-dependent mass term for some SYM scalars, 
$m^2(x)$ Tr$A^2(x)$, to 
the CFT Lagrangian, thereby explicitly breaking the global Poincare invariance 
(or conformal invariance for that matter). In the dual AdS  picture the 
perturbation is reflected by gravitating spacetime-dependent
fields turned on in the bulk, sourced from
the AdS boundary. 
But even away from the boundary, 
these fields break AdS isometries. This is a physical effect, apparent to a 
bulk observer.  Theoretically, one can phrase this as the turned-on fields
``Higgsing'' the bulk general coordinate invariance, although we usually do 
not adopt this language.
The second example is provided in Ref. \cite{kleb} which 
studied a supersymmetric field theory with a global $U(1)_R$ 
symmetry, explicitly broken by an anomaly. The field theory has a 
supergravity dual in which the global $U(1)_R$ is mapped to an AdS gauge 
field. The explicit breaking in the field theory must map to
 a Higgsing of the AdS gauge field. Therefore we know on general grounds that
the requisite Higgs condensate must 
appear on the AdS side of the dual, and Ref. \cite{kleb} 
shows in more detail that this is the case. 

~

Our final construction has the following properties:

~

(i) The deformation of SYM takes the form of a manifestly hermitian 
perturbation to the SYM Hamiltonian, thereby guaranteeing unitary and 
time-local quantum mechanical evolution.

~

(ii) The deformation explicitly violates the Poincare  
invariance of SYM. It mediates superluminal processes at first order.

~

(iii) To all orders in perturbation theory, the deformed theory is UV finite 
(no new divergences beyond the renormalized SYM CFT). The finiteness 
properties are related to the spatial non-locality
 of our perturbation Hamiltonian.
In particular, the deformed Hamiltonian is well-defined.

~

(iv) The deformed SYM theory is indeed a weak perturbation of all SYM
 processes, 
viewed for a finite period of time. That is, perturbation theory can be 
trusted. This implies that there must be some gravity dual of our deformed 
theory, including the dual of (ii).

~

(v) Some 
degree of superluminality in the gravity dual is taking place in the
AdS bulk, not just at the AdS boundary. Since the bulk spacetime 
gravitates, the necessary Higgs set-up must appear, but we are unable to 
give its explicit form.

~

(vi) Our deformed Hamiltonian is a sum of squares of 
hermitian operators and therefore has a energy bounded from below. But we 
cannot prove that the SYM vacuum is a ground state of the deformed theory. 
For instance, Ref. \cite{page} proves a positive energy theorem in the 
gravitational dual based on the assumption of relativistic causality of the 
CFT, but our deformation violates this assumption.
Instead, it is likely that the SYM vacuum
 corresponds to an excited state (not even 
an energy eigenstate), and therefore can decay into the true 
ground state of the full Hamiltonian. At first sight this appears 
problematic since we wish to consider propagation of simple objects in the 
``recognizable'' 
AdS vacuum, with weak superluminal corrections. The decay of the SYM vacuum 
implies the decay of the AdS vacuum into an unknown state. In such a state 
we would not necessarily know what spacetime metric to use to even define
superluminality. 

~

(vii) Fortunately, the vacuum decay rate can be controlled 
by the weakness of our perturbation,
 and we can allow the perturbation to be 
turned on for only a finite duration, thereby ensuring that most regions 
of space do not experience vacuum decay. This still leaves a small but 
non-zero amplitude for superluminal propagation and interaction of
 bulk quanta  in this (approximate) AdS(CFT) vacuum. 
In this way, we engineer rare breakdowns of the general relativistic 
approximation, with long-range superluminal consequences. 

~

The 
simultaneously 
weak and long-range character of the superluminal Lorentz-violating 
interactions
distinguishes our construction from 
earlier Lorentz-violating deformations of SYM  and their 
gravitational duals \cite{ganor}, 
and points to how such striking effects might 
be compatible with real world gravity.
  It is possible that the puzzling CFT interpretation 
of wormhole solutions in 
Euclidean AdS gravity \cite{rey} \cite{juan} \cite{nima}
is related to constructions similar in spirit to ours. However in
 this paper we work in Lorentzian signature spacetime. 
Several Lorentzian aspects of the AdS/CFT correspondence are discussed in 
Ref. \cite{adslor}.

We work up to our construction in the following stages. In Section 2, 
we illustrate the long-wavelength 
emergence of relativistic invariance, in the absence of 
gravity, in a simple quantum lattice model
which fundamentally has a preferred reference frame. 
We then add a perturbation that leads to
  instantaneous action-at-a-distance
in what would otherwise have been 
the continuum 
relativistic regime. In Section 3, we generalize this action-at-a-distance
 within a long-wavelength effective field theory containing gravity. 
The notion of ``instantaneous''  is ill-defined in a general relativistic 
context, but is replaced by superluminality. The requisite combination of 
Higgs effects is described. In Section 4, we review the emergent nature of 
(higher-dimensional) 
quantum general relativity, via the AdS/CFT correspondence, 
from ${\cal N} = 4$ SYM. We then describe the generalization of Section 2 to 
SYM. This section is primarily intended for contrast with Section 5, where 
our main results appear.
The AdS dual set-up contains superluminality in a general relativistic 
context, but localized to the $AdS_5$ boundary. The deformation also 
leads to UV divergences, that can however be treated by renormalization.
We explain 
how AdS/SYM 
vacuum decay can be suppressed by making the SYM deformation 
act for a finite duration. 
In Section 5, we 
describe a perturbation of the SYM Hamiltonian whose AdS dual contains 
superluminality in the AdS ``bulk''.
We show UV-finiteness and perturbativity 
 of our deformation to SYM, and again indicate how SYM vacuum decay can be 
suppressed by making the deformation act temporarily.
Section 6 provides our conclusions.


\section{Emergent Special Relativity and Action-at-a-distance}

Consider a very simple example.  We
  start with an underlying 
Hamiltonian for a theory without Poincare invariance  living on a 
spatial cubic lattice (continuous time),
\begin{equation}
H = \frac{1}{2}\sum_{\vec{n}} \{ \Pi_{\vec{n}}^2 +  \sum_{i=1}^3 
(\phi_{\vec{n} + \hat{i}} - \phi_{\vec{n}})^2 \},
\end{equation}
where $\hat{i}$ are spatial unit vectors, and $\vec{n}$ are lattice points.
In familiar fashion, for long wavelength modes of this system we arrive at the 
approximately relativistic theory of a massless free scalar field,
\begin{equation}
H \approx \frac{1}{2} 
\int d^3 \vec{x} \{ \Pi^2(\vec{x}) + (\partial_i \phi(\vec{x}))^2 \}.
\end{equation}
Quantization of both Hamiltonians, written above in terms of 
Schrodinger picture operators, is of course straightforward.
Even in this simplest of examples, the underlying theory contains 
couplings, $\phi_{\vec{n} + \hat{i}} \phi_{\vec{n}}$, which 
instantaneously connect two points at finite spatial separations.

But we can arrange for a more drastic breakdown of Poincare invariance 
right in the midst of the relativistic regime, for example,
\begin{eqnarray}
\label{scalar}
H &=& \frac{1}{2} \sum_{\vec{n}} \{ \Pi^2_{\vec{n}} +  \sum_{i=1}^3 
(\phi_{\vec{n} + \hat{i}} - \phi_{\vec{n}})^2 \} \nonumber \\
&+& \epsilon (\sum_{\vec{n}} J_{\vec{n}} \phi_{\vec{n}})^2,
\end{eqnarray}
where $J \geq 0$ is a lattice function with finite support, and normalized to 
\begin{equation}
\sum_{\vec{n}} J_{\vec{n}} = 1. 
\end{equation}
The entire 
perturbation to the Hamiltonian 
is obviously also $\geq 0$ and minimized at the 
vacuum $\langle \phi \rangle = 0$. 
If the support of $J$ has a typical size $L \gg$ lattice-spacing $\equiv 1$ 
and $J$ is smooth on that scale, then the perturbation can be made weak by 
taking $\epsilon \ll L$. In the support region of $J$, 
 a lattice quantum can be absorbed by the perturbation in one location 
$\vec{n}_1$ and be
 instantaneously emitted at a distant location $\vec{n}_2$.

Note that because energy is conserved and because 
 $J$ is smooth on the lattice scale, 
soft incoming quanta necessarily scatter (non-locally in space) 
to soft quanta, so that 
the continuum long-wavelength approximation is not broken by the perturbation. 
There is therefore a good continuum approximation to this model,
\begin{eqnarray}
\label{target}
H &\approx& \frac{1}{2} 
\int d^3 \vec{x} \{ \Pi^2(\vec{x}) + (\partial_i \phi(\vec{x}))^2 \}
\nonumber \\
&+& \epsilon (\int d^3 \vec{x} J(\vec{x}) \phi(\vec{x}))^2, 
\end{eqnarray}
where $J$ is normalized as
\begin{equation}
\int d^3 \vec{x} J(\vec{x}) = 1.
\end{equation}
Again, the weak $\epsilon$ coupling can absorb 
relativistic quanta and instantaneously emits such quanta far away. Such 
effects can 
appear acausal in another relativistic reference frame, but that frame is not
co-equal to the defining  one above. 
We see that the lattice structure is irrelevant to the 
question of this type of long-range interaction. Once one declares 
there to be a  preferred frame, all that is required is a causal unitary 
theory in that frame. It may have a relativistic approximation when 
some (in this case, long-range) interactions are neglected, but this 
relativity is not an exact principle that disqualifies
 non-relativistic perturbations.

While the original lattice model, being just a discrete set of quantum
 mechanical degrees of freedom is manifestly UV-finite, we should check that 
this is the case for the continuum approximation (that is check that we can 
 truly decouple the lattice structure and have a continuum limit). Thinking 
of $\epsilon$ as a perturbation quadratic in fields, we see that the only 
(potentially divergent) loop diagrams are to vacuum energy. Since there is 
no gravity in this model, vacuum energy is physically irrelevant and we 
can ignore these diagrams. 

\section{Effective Gravity and Superluminality}

Are superluminal long-range effects consistent with General Relativity? 
Let us try to construct
 a long-wavelength continuum description  of such a combination,
 generalizing (\ref{target}). In this section we will not worry about 
the issue of UV completeness.
To simplify our task a little, let us first aim for the limit in which $J$ 
is supported on just two points between which we want to arrange for 
superluminal interactions, 
  $J(\vec{x}) = \frac{1}{2} \delta^3(\vec{x} - \vec{x}_{1}) + 
\frac{1}{2} \delta^3(\vec{x} - \vec{x}_{2})$, temporarily 
turning a blind eye to the loss of smoothness and the product of 
coincident $\delta$-functions. We will add smoother sources at the end of 
this section.
Since spacetime 
symmetries are gauged by General Relativity, we will realize the 
breaking of Poincare invariance as a Higgs effect. 
Really, two Higgs effects are required: one to pick out the two 
special locations, $\vec{x}_1, \vec{x}_2$, and one to define 
``simultaneous'' times on these two locations at which the long-range 
interactions occur. 

We take the first Higgs effect to be of a familiar type:
we add to our theory a new species of particle, $\psi$, with mass $m$ much 
greater than the UV cutoff of our effective description, but smaller than 
the Planck scale. We also assign it a ${\bf Z}_2$ 
charge so that it can only be destroyed or created
 in pairs. 
Heavy pairs cannot be created within 
the effective description from 
light gravitational and $\phi$ quanta much softer than $m$, but the light 
particles can interact with a pre-existing $\psi$ pair. We take this pair of
heavy particles to be so distantly separated at some initial time, that 
they cannot annihilate for a very long time to come. Since light and soft 
quanta cannot appreciably accelerate the massive $\psi$, we can take 
the $\psi$ pair to be approximately at rest with respect to an asymptotic 
Minkowski frame. Because of their large inertia and small Compton wavelength 
with respect to the UV cutoff of the light quanta, 
the pre-existing $\psi$ particles will act as 
effectively fixed pointlike locations. By this means, the ``preferred'' 
locations of 
the $\psi$-pair effectively spontaneously breaks (local) Poincare 
invariance.
Their two locations will generalize 
the fixed locations $\vec{x}_{1,2}$ of our long-range interaction. 

We want an interaction between $\psi$ and $\phi$ so that a 
$\phi$ quantum that propagates to the location of one of the pre-existing 
$\psi$'s can ``instantly'' jump to the location of the other $\psi$. 
But this requires an identification of time on the worldline of one of the 
$\psi$'s with time on the worldline of the other distant $\psi$.
Such a preferred pairing of times  further 
breaks general coordinate invariance 
and necessitates the second Higgs effect. Minimally, this Higgs effect 
can be localised to the $\psi$ worldlines, defining the preferred times as 
the proper time along each $\psi$ worldline since their pair-creation in the 
distant past. However, it is convenient to use a Higgs effect
already in the literature that defines a 
preferred time everywhere in space, namely the ``Ghost-condensate'' 
\cite{ghost}. In a generally covariant and consistent, but unusual, effective 
field theory a  new scalar field, $\chi$,
 is coupled to gravity so as to admit a non-trivial stable 
solution, namely Minkowski 
spacetime metric with 
\begin{equation}
\label{ghost}
\chi(x) = k x^0,
\end{equation}
 where $k$ is a fixed constant parameter from the $\chi$ action. 
This time dependence 
arises spontaneously and partially Higgses general coordinate invariance. 
Small  $\chi$ flucuations about (\ref{ghost})
are ``eaten'' by the metric fields. Therefore in unitary gauge (\ref{ghost}) is exact 
while the gravitational action is modified. Nevertheless, over a large regime 
this effective field theory reproduces standard General Relativity. 
The field (\ref{ghost}) then gives us a global ``clock''. 

Putting together the ingredients, we
 take our model of superluminality to be given by
\begin{eqnarray}
S &=& S_{Einstein}[g_{\mu \nu}] + S_{ghost}[g_{\mu \nu}, \chi] 
+ \frac{1}{2} 
\int d^4 x \sqrt{-g} \{g^{\mu \nu}
\partial_{\mu} \phi \partial_{\nu} \phi  + 
g^{\mu \nu}
\partial_{\mu} \psi \partial_{\nu} \psi - m^2 \psi^2 \} \nonumber \\
&-& \frac{\epsilon}{4} \int d^4 y \sqrt{-g(y)} \psi^2(y) \phi(y) \int d^4 z 
\sqrt{-g(z)} \psi^2(z) \phi(z) 
~ k~ \delta(\chi(y) - \chi(z)).
\end{eqnarray}
First note that this action is generally coordinate invariant. 
Somewhat similar non-local operators were discussed in Ref. \cite{giddings} 
as coordinate invariant observables in ordinary effective general relativity.
Here, the non-local operator represents a true 
modification of the dynamics, not 
just a probe of standard gravity. 

After 
passing to the ghost condensate unitary gauge the non-local term above 
becomes
\begin{equation}
S_{superluminal} = - \frac{\epsilon}{4} \int dt
\int d^3 \vec{y}  \sqrt{-g(t, \vec{y})} \psi^2(t, \vec{y}) \phi(t, \vec{y}) 
\int d^3 \vec{z} \sqrt{-g(t, \vec{z})}
\psi^2(t, \vec{z}) \phi(t, \vec{z}),
\end{equation}
which is non-local in space, but local in time. The leading behavior of this 
system can be seen in the limit that the UV cutoff is $\ll m \ll M_{Pl}$. 
In the limit, with $k$ held fixed, gravity decouples from the dynamics, but 
$\chi$ continues to provide a global time. 
 The pair of distant heavy $\psi$ particles become infinitely massive and 
point-like, with static locations $\vec{x}_{1,2}$. Therefore $\phi$ 
effectively has an action in this limit,
\begin{equation}
S_{eff} = \frac{1}{2} \int d^4 x (\partial_{\mu} \phi)^2 - \frac{\epsilon}{4} 
\int dt (\phi(t, \vec{x}_1) + \phi(t, \vec{x}_2))^2,
\end{equation}
which is equivalent to (\ref{target}) with $J = \frac{1}{2} \delta^3 
(\vec{x} - \vec{x}_{1}) + \frac{1}{2} \delta^3 
(\vec{x} - \vec{x}_{2})$. 

With $M_{Pl}, m$ large but finite, the 
$\epsilon$ interaction is instantaneous in the unitary gauge, but not in a 
fully general coordinate invariant sense. Rather, the general statement is 
that  the $\epsilon$ interaction is {\it superluminal} with respect 
to the metric $g_{\mu \nu}$. 

Finally, let us discuss how to generalize this construction to allow 
a smooth $J(\vec{x})$. The simplest way is to replace the 
pair of heavy elementary particles $\psi$ with a  smooth soliton and an
anti-soliton. For example suppose $\psi^{a}$ is an isotriplet Higgs field  
for an $SO(3)$ gauge theory, which Higgses the symmetry down to $SO(2)$ 
(Georgi-Glashow model \cite{gg}).
This theory supports smooth magnetic  monopole solitons.  
We take our heavy pre-existing particles to be a distantly separated 
monopole $+$ anti-monopole pair. Let us generalize 
our $\psi^2 \phi$ couplings above to $(\psi^a \psi^a - v^2) 
\phi$, where $v$ is the magnitude of the $\psi^a$ VEV. 
Therefore the $\epsilon$ interaction turns on smoothly as one enters the cores 
of the monopoles where $\psi^a$ deviates appreciably from its VEV.
The $J$ we have engineered has negligible support except in the two widely 
separated soliton cores.

It is not known if this relatively simple 
effective gravitational dynamics exhibiting superluminality can be UV 
completed, but it is a sensible low-energy effective field theory (at least 
over a finite but long time interval to avoid any gravitationally induced 
collapse) and 
illustrates the principles we will pursue, indirectly,
 in the context of AdS/CFT.

\section{AdS/CFT and Boundary Superluminality}  

The $CFT \rightarrow AdS$ correspondence is in a very real 
sense a case of emergent gravity and relativity. 
The UV completeness of the CFT transfers to the AdS quantum gravity.
Let us specialize to 
the CFT given by strongly-coupled large-$N_{color}$ ${\cal N} = 4$ SYM. 
There are good arguments \cite{kaplan} \cite{simeon} \cite{strassler}
to suggest that this theory might itself 
be realizable as the IR limit 
of a  lattice theory (continuous time) with a preferred frame for 
unitary quantum evolution. 
Such a lattice system would 
violate all of Poincare invariance except for time translation invariance. 
Poincare and conformal invariance would emerge in the continuum  
long-wavelength limit. The gravity dual of  such a lattice system would have 
a ``UV'' boundary at which Poincare invariance is badly broken, reflecting 
the YM lattice structure,
with IIB superstring field profiles emanating from the UV boundary,
 perturbing the usual $AdS_5 \times S^5$ 
background.
But the dual of the statement that the far IR of the lattice theory is 
successfully approximated by 
continuum SYM translates to saying that, 
far away from the UV boundary in the
IR of the bulk, the  $AdS_5 \times S^5$ background and fluctuations are
 gradually restored (the deviating profiles damp out).
In this sense, (higher dimensional) 
General Relativity can emerge from a quantum theory which fundamentally does 
not enjoy (even special) relativistic structure. The above features follow 
on general grounds, but details of the
 AdS dual of such a lattice gauge theory are not known. However, a provocative 
related example, with a single lattice dimension,
 has been studied in Ref. \cite{simeon}. A general moral to keep in mind 
is this: 
if a UV complete quantum theory has a regime or approximation in which it 
matches a CFT which has an AdS gravity (string) 
dual, then the entire quantum theory 
must have a dual description which has a gravitational 
regime or approximation. This latter gravitational (string) 
dual must also reflect 
the deviations from  CFT behavior, and  must possess the 
objects and defects necessary to do so. 

For most of  this section
we work directly in the continuum (only briefly 
invoking a possible 
lattice realization in subsection 4.4). 
We generalize
(\ref{target}) by perturbing the SYM CFT by a bilocal interaction, 
but now each local factor must be a $SU(N_{color})$ gauge-invariant composite 
operator. SYM has six ``flavors'' of 
real color-adjoint scalar fields, $A_{I = 1,..6}$. 
Flavor-adjoint color-singlet scalar bilinears, 
Tr$A_I A_J$ - $\frac{\delta_{IJ}}{6}$ Tr$A_K A_K$, are primary operators of the
SYM CFT of dimension 2 (related by extended supersymmetry to conserved 
currents).  We 
will pick any of them, say ${\cal O}(x) \equiv$ Tr$A_1 A_2$, to build a
bilocal perturbation to SYM:
\begin{equation}
\label{Hbd}
H = H_{CFT} + \epsilon (\int d^3 \vec{x} J(\vec{x}) {\cal O}(\vec{x}))^2. 
\end{equation}
Here the operator ${\cal O}$ is in Schrodinger picture.
 We have chosen a very 
low-dimension operator 
so as to minimize the issues of UV divergences, studied in subsection 4.2. 
Local double-trace operator deformations were 
studied in Refs. \cite{eva} \cite{witten} \cite{berkooz} \cite{minces} 
\cite{ofer}. 

We can also pass to the action 
formulation and  path integral quantization:
\begin{eqnarray}
\label{action}
S &=& S_{CFT} -  \epsilon \int dt (\int d^3 \vec{x} J(\vec{x}) 
{\cal O}(t, \vec{x}))^2 \nonumber \\
 &=& S_{CFT} - \epsilon \int d^4 x J(\vec{x}) {\cal O}(x) 
\int d^4 y J(\vec{y}) {\cal O}(y) \delta (x_0 - y_0). 
\end{eqnarray}

\subsection{Superluminality}

We consider the reference frame of $H$ as the preferred one in which 
quantum time-evolution is defined.  
As in Section 2, 
the $\epsilon$ perturbation
 can absorb CFT excitations and instantly re-emit them far away in the 
support of $J$, consistent with causality in the defining frame. 
This effect is reflected as superluminality in the gravity dual. 
Without $\epsilon$, the dual vacuum configuration is of course 
the well known $AdS_5 \times S^5$. For the point we want to make, the 
$S^5$ is just a detail. We will not bother keeping track of locality on the
$S^5$, just Kaluza-Klein reducing from 10 dimensions down to 5. 
Choose Poincare coordinates in AdS, in the 
same preferred frame as the perturbed CFT,
\begin{equation}
\label{ads}
ds^2_{AdS} = \frac{\eta_{\mu \nu} d x^{\mu} d x^{\nu} - d z^2}{z^2}. 
\end{equation}

Let us focus on the propagation of the
 AdS scalar, $\phi(x,z)$, 
dual to the operator ${\cal O}$. It is a ``good'' tachyon with 
5D mass-squared of $-4$, saturating the Breitenlohner-Freedman stability 
bound \cite{bf}. 
Consider two spacelike-separated 
events in the AdS 
bulk spacetime, $(0, \vec{0}, z)$ and $(t > 0, \vec{x}, 
z)$, with $2z < t \ll |\vec{x}|$,
so that causal communication between them
is ordinarily ($\epsilon = 0$) impossible.  
 However, let us now suppose that $\vec{0}$ and 
$\vec{x}$ are both within the support of $J$. To first order in $\epsilon$, 
perturbation theory pulls down from the action
$\epsilon \int d t' (\int d^3 \vec{x}' J(\vec{x}') {\cal O}(t', \vec{x}'))^2$.
The AdS 
dual of this perturbation at leading order in large $N_{color}$ 
is that each ${\cal O}(t', \vec{x}')$ 
maps to a bulk-boundary 
free-field AdS propagation of the $\phi$ scalar, 
with the boundary point being 
$(t', \vec{x}', 0)$. Denoting the free-field bulk-boundary propagator between 
$(x,z)$ and $x'$
by $K(x-x',z)$, we see that our leading correction to the bulk-to-bulk 
propagator
 between $(0, \vec{0}, z)$ and $(t > 0, \vec{x}, 
z)$ is 
\begin{equation}
\label{K}
\epsilon \int dt' \int d^3 \vec{x}' d^3 \vec{y}' J(\vec{x}') 
J(\vec{y}') K(t',\vec{x}',z) K(t-t', \vec{x} - \vec{y}', z).
\end{equation} 
Now for example, boundary points such as $(t' \approx t/2, \vec{x}' \approx \vec{0})$ and 
$(t' \approx t/2, \vec{y}' \approx \vec{x})$ contribute to this integral. 
In AdS, such boundary points are causally connected to our bulk points 
$(0, \vec{0}, z)$ and $(t > 0, \vec{x}, 
z)$ respectively. That is each $K$ factor allows
 causal communication (they have 
imaginary parts) and hence so does the entire bulk-to-bulk correction. 
This result is also a limiting case of that of Section 5, which 
provides a more formal derivation.

(This leading order superluminality is UV-finite in the continuum. Therefore 
it is also a good approximation to a lattice realization of the deformed 
SYM theory, as long as all the relevant length scales are much larger 
than the lattice spacing, in particular 
$t, |\vec{x}|, z$ above and the dominant wavelengths of $J$.)

 A quantum gravitational theory thereby admits superluminal 
propagation, although in this case the ``magic'' is localized to the AdS 
boundary. (A brief comment to similar effect is made in the discussion of 
Ref. \cite{nima}.)
Still, the AdS string theory must possess the necessary boundary 
defects that allow this to occur, as long as our
CFT deformation is UV complete. But we cannot argue that this takes the form of
a Higgs mechanism on 
the AdS boundary since in a sense gravity and general coordinate invariance 
end there.

\subsection{Renormalizability}

We must take care to understand what divergences emerge due to the 
multiple operator products of ${\cal O}$ appear in 
$\epsilon$ perturbation theory. The reader may wish to follow this section 
by using the free SYM field theory as a simple example. Although we are 
primarily 
interested in strongly coupled SYM so that the AdS dual is weakly coupled, 
the operator 
${\cal O}$ has very similar divergence properties at arbitrary  coupling 
because of its supersymmetry-protected dimension.

 We will power-count $\epsilon$ perturbation theory 
 to identify the superficial divergences. 
The bi-local nature of our perturbation makes 
this somewhat unfamiliar. We can massage it a little to make it more amenable 
to standard power-counting. 
We attribute our perturbation to one that is linear in ${\cal O}$,
\begin{equation}
\label{fake}
\Delta S = \int d^4 x \epsilon^{1/2} J(\vec{x}) {\cal O}(x) \sigma(x), 
\end{equation}
where $\sigma$  is an auxiliary field with a ``propagator'', for 
$\epsilon =0$,
\begin{equation}
G_0(x,y) \equiv - 2i \delta(x_0 - y_0).
\end{equation}
We can use this propagator to
 ``integrate out'' $\sigma$ and return to the original 
perturbation. This is a formal device in that $G_0$ does not follow from 
some quadratic $\sigma$ action, but it is useful for power-counting 
purposes.
Since ${\cal O}$ has scale dimension $2$,  for 
power-counting purposes the background field 
$\epsilon^{1/2} J(\vec{x})$ has dimension $3/2$, 
and from its propagator it is clear that 
 $\sigma(x)$ has power-counting dimension $1/2$. 

Let us consider what superficial divergences there can be in the 
$\epsilon$ perturbation expansion. 
These must be local products of CFT operators 
multiplied by powers 
of $\sigma(x)$ and $\epsilon^{1/2} J$ (and derivatives), 
with total dimension $\leq 4$. 
Since a single operator 
insertion of ${\cal O}$ in the CFT is not divergent, the divergences 
can only begin at quadratic order in $\epsilon^{1/2} J$, already ``costing'' 
dimension 3. At most this could 
be multiplied by powers of $\sigma$, since all CFT gauge invariant operators
have scaling dimension $>1$. By $\sigma \rightarrow - \sigma, J \rightarrow -J$
symmetry, the only 
divergent structures can be $\epsilon J^2(\vec{x})$ and  
$\epsilon J^2(\vec{x}) \sigma^2(x)$. The first of these divergent 
structures can only 
 arise from the leading order VEV of the perturbation Hamiltonian, a 
physically irrelevant real c-number constant,
 that can be renormalized away by simply
subtracting it from our Hamiltonian (or action).

We are thus only left to contend with  
$\epsilon J^2(\vec{x}) \sigma^2(x)$, that is, a divergence in the 
$\sigma$ self-energy.
Indeed there really is a logarithmic 
divergence of this form, and it is coupled to the 
rest of the CFT because each $\sigma(x)$ field in it can be contracted 
with $\sigma$'s elsewhere in the perturbative expansion, 
so we do have to address this divergence. The leading correction to the 
$\sigma$ self-energy takes the form
\begin{eqnarray}
\label{ope}
&~& \epsilon \int d^4 x J(\vec{x}) \sigma(x) 
\int d^4 y J(\vec{y})  \sigma(y) \langle 0|T \{{\cal O}(x) {\cal O}(y) \} |0
\rangle \nonumber \\
&~& ~ ~ ~ ~ ~ 
= \epsilon c \ln (\mu a) \int d^4 x J^2 (\vec{x}) \sigma^2(x) {\rm +
finite},
\end{eqnarray}
where $a$ is a short-distance cutoff and $\mu$ is an arbitrary renormalization 
scale put in to separate out the UV divergence, 
and $c$ is an unimportant constant. We have used that ${\cal O}$ is a 
dimension 2 primary operator
 in constraining the form of its correlator in the usual way.

This divergence is removed by renormalization of $\epsilon$. To see this 
let us resum the $\sigma$ self-energy contributions to its propagator 
arising from integrating out the CFT. Define the self-energy correction
\begin{equation}
\epsilon \Pi(x,y) = \epsilon J(\vec{x}) J(\vec{y}) \langle 0| T \{ {\cal O}(x) 
{\cal O}(y) \} |0 \rangle.
\end{equation}
Log divergences are removed by the subtraction
\begin{equation}
\epsilon \Pi_{sub}(x,y) = \epsilon \Pi(x,y ) - \epsilon \Pi_{div}(x,y), 
\end{equation}
where by (\ref{ope})
\begin{equation}
\epsilon \Pi_{div}(x,y) \equiv \epsilon c
\ln(\mu a) J^2(\vec{x}) \delta^4(x-y).
\end{equation}

Resumming this self-energy in the $\sigma$ propagator gives, 
in an obvious matrix notation,
\begin{eqnarray}
\epsilon G &=& \epsilon G_0 (I - \epsilon \Pi G_0)^{-1} \nonumber \\
&=& \epsilon G_0 (I - \epsilon \Pi_{sub} G_0 - \epsilon \Pi_{div} G_0)^{-1}.
\end{eqnarray}
This is the only combination in which $\epsilon$ and $G_0$ appear in the 
perturbative expansion, and all powers of $\epsilon$ are explicitly shown.
Note that in the expansion of this expression $\Pi_{div}$ always appears 
in the sandwich $G_0 \Pi_{div} G_0$, for which it is easy to prove,
\begin{equation}
G_0 \Pi_{div} G_0 = \ln(\mu a) {\cal J} G_0,
\end{equation} 
where ${\cal J}$ is just the finite constant 
\begin{equation}
{\cal J} \equiv c \int d^3 \vec{x} J^2(\vec{x}).
\end{equation} 
Using this relation, we can rewrite the resummed $\sigma$ propagator as
\begin{eqnarray}
\epsilon G &=& \epsilon G_0 
(I - \epsilon \Pi_{sub} G_0 - \epsilon {\cal J} \ln(\mu a) I)^{-1} \nonumber \\
&=& \epsilon_R(\mu) G_0 (I - \epsilon_R(\mu) \Pi_{sub} G_0)^{-1},
\end{eqnarray}
where we define a renormalized coupling at $\mu$,
\begin{equation}
\epsilon_R(\mu) \equiv \frac{\epsilon}{1 - \epsilon {\cal J} \ln(\mu a)}.
\end{equation}

Since there are no more superficial divergences other than the real 
divergence subtracted by 
this renormalization, the perturbative expansion is now finite in terms of 
$\epsilon_R$ as the short-distance cutoff $a \rightarrow 0$.

\subsection{Suppressed Vacuum decay}

As discussed in the Introduction, there is no guarantee that the SYM vacuum 
remains the true ground state of the deformed theory. However, whatever 
the true ground state, the amplitudes for the decay of the SYM 
vacuum follow from our renormalizable theory, giving some finite 
decay rates (per unit volume) perturbatively in $\epsilon_R$. Above, we did 
subtract a single infinite (order $\epsilon J^2$) 
correction to vacuum energy as part 
of renormalization, but this is irrelevant for vacuum decay since 
this divergence is real while it is the imaginary parts of ``vacuum''-energy
that encode vacuum decays via the optical theorem.
For sufficiently small
$\epsilon_R$ (renormalized at the scale typical of $J$) 
these decay rates, and Lorentz-violating processes in general, 
are suppressed, but a sufficiently long time will always overcome the weak 
coupling and lead to complete
SYM(AdS) decay. To prevent this from happening 
we will consider $\epsilon(t)$ to be smoothly 
time-dependent with finite support. This makes the Hamiltonian time-dependent.
Now we can choose $\epsilon_R$ so weak as to not lead to catastrophic 
vacuum decay. This also suppresses the probability of superluminal 
propagation, but it does not vanish and we are only seeking this 
qualitative fact. Our previous analysis of superluminality in subsection 
4.1 is only altered in that the $\epsilon \rightarrow \epsilon(t')$ now 
sits inside the time integral. Superluminality continues to hold as long 
as the duration of non-vanishing $\epsilon$ is taken to cover
 the events being discussed there.



\subsection{Comments}

We have seen that the deformed CFT is renormalizable in the same sense that 
QED is, 
logarithmic UV divergences being eliminated by $\epsilon$ renormalization,
but the catch is that $\epsilon$ runs in the UV to strong coupling where our 
power-counting breaks down. Thus, we have not yet demonstrated true 
UV completeness, but we are close.
Of course, one possibility is that 
the short-distance cutoff $a$ may really be finite, if for example 
it is a lattice spacing for SYM on a spatial lattice. 
A spatial lattice theory would
regulate the divergences of the continuum 
field theory by converting it
into quantum mechanics of discrete lattice degrees of freedom. 
In that case, the renormalizability usefully translates into insensitivity of 
the long-wavelength theory to details of the lattice structure. 

Another possibility is not to regulate the CFT itself, but rather to 
replace each ${\cal O}(x)$ with a ``slightly'' 
 non-local operator so as to regulate the 
operator product 
divergences appearing in $\epsilon$ perturbation theory. A seemingly separate 
question is whether superluminality can be realized in the gravitating 
bulk of AdS, rather than the AdS boundary where the gravitational dynamics 
ends. We really would like to test if superluminality can appear right in 
the midst of quantum General Relativity. Presently, bulk quanta must propagate
relativistically to the AdS boundary before they see superluminal effects. 
As it turns out, we can indeed engineer bulk superluminality, and 
the trick is to replace the pair of local operators appearing in the 
CFT Hamiltonian by a pair of non-local operators. As an added benefit, 
this replacement renders the deformed theory UV finite, that is, it 
regulates the UV divergences discussed above. We pursue this approach next.

\section{Superluminality in the Bulk}


In this section,  ${\cal O}$ 
denotes an arbitrary local scalar primary operator of the SYM CFT, of scaling 
dimension $d$, which has single-trace limit as $N_{color} \rightarrow \infty$.
In subsection 5.4 we will restrict $d$ to ensure perturbativity of our 
deformation under all circumstances. Until then we simply assume 
perturbation theory in the CFT deformation is to be trusted.

\subsection{Warm-up on non-local operators and UV finiteness}

 Consider the simple, but time-dependent, 
deformation of the CFT Hamiltonian of the form
\begin{eqnarray}
H(\tau) &\equiv& H_{CFT} + \Delta H(\tau) \nonumber \\
&\equiv& H_{CFT} +
\int d^3 \vec{x} J(\tau, \vec{x}) {\cal O}(\vec{x}).
\end{eqnarray}
All the operators appearing here are Schrodinger operators, 
time-dependence appearing only in the source, $J(\tau, \vec{x})$.

Since time-ordering subtleties will be important in this section, we 
use Hamiltonian and operator methods throughout. The master formula for 
time-ordered perturbation theory is
\begin{equation}
\label{pert}
T e^{- i \int_{t_1}^{t_2} d \tau H(\tau)} = e^{- i H_{CFT} (t_2 - t_1)} 
\sum_{n=0}^{\infty} 
\frac{(-i)^n}{n !} \int_{t_1}^{t_2} d \tau_1 \int_{\tau_1}^{t_2} d \tau_2 ...
\int^{t_2}_{\tau_{n-1}} 
d \tau_n \Delta \hat{H}(\tau_n) ... \Delta \hat{H}(\tau_1).
 \end{equation}
We uniformly use hats to distinguish Heisenberg operators,
\begin{eqnarray}
\Delta \hat{H}(\tau) &\equiv& e^{i H_{CFT} \tau} \Delta H(\tau) 
e^{-i H_{CFT} \tau} \nonumber \\
\hat{\cal O}(\tau, \vec{x}) &\equiv& e^{i H_{CFT} \tau} {\cal O}(\vec{x})
e^{-i H_{CFT} \tau}.
\end{eqnarray}

Consider the example of the CFT-vacuum persistence amplitude at order $J^2$,
\begin{eqnarray}
&~& \int d^4 x d^4 y J(x) J(y) 
\langle 0 | T \{ \hat{{\cal O}}(x) \hat{{\cal O}}(y) \} |0\rangle 
\nonumber \\
&=& 
\int d^4 x d^4 y J(x) J(y) 
\int dm^2 \int d^3 \vec{p} 
\frac{|\langle 0| {\cal O}(0) 
|m, \vec{p} \rangle|^2}{2 \sqrt{\vec{p}^2 + m^2}}  
(\theta(x_0 -y_0) 
e^{-ip.(x-y)} + \theta(y_0 -x_0) e^{ip.(x-y)}) \nonumber \\
&\propto& \int d^4 x d^4 y J(x) J(y) 
\int dm^2 m^{2 d -4} G(x-y;m) \nonumber \\
&=& \int d m^2 m^{2 d -4} \int d^4 q \frac{|\tilde{J}(q)|^2}{q^2 - m^2 + 
i \epsilon} \nonumber \\
&=& \infty.
\end{eqnarray}
The first equality follows in passing to the 
 spectral representation by inserting a complete set of 
states, integrating over their invariant mass-squared and momentum. The 
proportionality follows from the fact that by Lorentz invariance 
the matrix element of the scalar 
operator ${\cal O}$ can only depend on the mass $m$, not the momentum 
$\vec{p}$, and the mass-dependence follows by dimensional analysis in the 
CFT. $G(x-y;m)$ denotes a free field scalar Feynman propagator of mass $m$.
The final equality arises because, whereas the smoothing due to $J$ 
cuts off the $q$-integral, the $m$ integral remains divergent. 

This divergence is very closely related to the one of the previous section, 
and is rather standard when perturbing by (superpositions of) 
local 
operators. In perturbation theory, $\hat{\cal O}(\tau, \vec{x})$ 
creates a pointlike disturbance at time $\tau$, which then begins to 
spread out. However, at second order a second $\hat{\cal O}(\tau', \vec{x}')$ 
can 
sample the disturbance created by the first, and this creates our divergence 
when the two points coincide. Such divergences are avoided if the pointlike 
disturbances are ``thickened'' to finite size. One convenient way of 
doing this is to use a fake time evolution to spread out the pointlike 
disturbance created by ${\cal O}$.  A simple illustration is provided by the 
Hamiltonian, 
\begin{equation}
H = H_{CFT} + \int d^4 x J(x) e^{i H_{CFT} x_0} {\cal O}(\vec{x}) 
 e^{-i H_{CFT} x_0}.
\end{equation}
All the operators appearing are in Schrodinger picture. One product
of these operators happens to be a Heisenberg operator in form, but the whole
 effect of the $x_0$ ``time evolution'' on ${\cal O}$ is to 
turn this spatially local Schrodinger operator into a spatially non-local 
one, by evolving the disturbance it creates  for a finite time which 
causes the disturbance to spread over a finite spatial region.  Note that 
since $x_0$ is integrated over, the Schrodinger $H$ and $\Delta H$ are 
time-independent. It is important to note that by specifying a Hamiltonian, 
the resulting dynamics is automatically local in time. But the perturbation is 
spatially non-local because it cannot be written as a superposition of 
local Schrodinger operators. If this looks unfamiliar it is because 
it is inconsistent with Lorentz invariance, from which we are deviating 
in this paper.

To contrast this with our earlier example, let us again apply (\ref{pert}) 
to calculate
the CFT-vacuum persistence amplitude at order $J^2$, 
\begin{eqnarray}
&\propto& \int_{- \infty}^{\infty} d \tau \int d^4 x d^4 y J(x) J(y) 
\langle 0|T_{\tau} \{ \hat{\cal O}(x_0 + \tau, \vec{x}) \hat{\cal O}(y) \}
 |0 \rangle 
\nonumber \\
&=& 
  \int d \tau \int d m^2 \int \frac{d^3 \vec{p}}{2 \sqrt{\vec{p}^2 
+ m^2}} \{ \theta(\tau) e^{-i  \sqrt{\vec{p}^2  + m^2} \tau} 
+ \theta(-\tau) e^{i  \sqrt{\vec{p}^2 +m^2} \tau} \} \nonumber \\
&~& ~ ~ ~ ~ ~ ~ ~ \times 
 |\tilde{J}(\sqrt{\vec{p}^2 + m^2}, \vec{p}) 
\langle 0| \hat{\cal O}(0) |m, \vec{p} \rangle|^2 \nonumber \\
&\propto& 
\int d \tau \int d m^2 m^{2d-4} 
\int \frac{d^3 \vec{p}}{2 \sqrt{\vec{p}^2 
+ m^2}} |\tilde{J}(\sqrt{\vec{p}^2 + m^2}, 
\vec{p})|^2 \nonumber \\
&~& ~ ~ ~ ~ ~ ~ ~ \times
\{ \theta(\tau) e^{-i  \sqrt{\vec{p}^2  + m^2} \tau} +
\theta(-\tau) e^{i  \sqrt{\vec{p}^2 + m^2} \tau} \}  \nonumber \\
&<& \infty.
\end{eqnarray}
The time-ordering is with respect to $\tau$ only.
Here, we see that the smoothness of $J$ does cut off   the $m$ and $\vec{p}$ 
integrals, and the $\tau$ integral also converges as $\tau \rightarrow 0$. 
This merely reflects the ``thickened'', 
as opposed to completely local, operator
that perturbs the CFT in the second example.


The comparison of the relatively simple examples in this subsection should 
orient the reader in the full construction of the next.

\subsection{Superluminality}

Consider the Hamiltonian,
\begin{eqnarray}
H &=& H_{CFT} + \Delta H \nonumber \\ 
&\equiv& 
H_{CFT} + \epsilon (\int d^4 x J(x) e^{iH_{CFT} x_0} {\cal O}(\vec{x})
e^{-iH_{CFT} x_0})^2,
\end{eqnarray}
where everything has been written in terms of Schrodinger operators, and 
where $J$ is  a smooth 
spacetime-dependent 
source of compact support. Note that $x_0$ is  a dummy integration 
variable and that $H$ is in fact time-independent. (We will however 
introduce time-dependence in subsection 5.4.)

To demonstrate superluminality, we consider properties of 
the bulk-to-bulk propagator
 for the scalar 
$\phi$,  dual to the primary operator ${\cal O}$, between the two spacetime 
points $(y^{\mu},z)$ and $(0, z)$, 
namely $\langle 0| T \{ \hat{\phi}(y,z) 
\hat{\phi}(0,z) \} 
|0\rangle$. (This a well-defined object 
if we imagine having  gauge-fixed general coordinate invariance.) 
Our approach (but not our result) 
shares some similarities with that of Ref. \cite{ofer} studying 
AdS implications of double-trace but local deformations of SYM.
We will take 
\begin{equation}
0 < y_0 < |\vec{y}|, 
\end{equation}
so that the two bulk points are ordinarily ($\epsilon =0$) 
causally disconnected (recalling that our AdS metric is (\ref{ads})), 
implying the vanishing of the commutator,
\begin{equation}
\label{order}
\langle 0| [\hat{\phi}(y,z), \hat{\phi}(0,z)] |0\rangle = 
2i ~{\rm Im} \langle 0| T \{ \hat{\phi}(y,z) \hat{\phi}(0,z) \} 
|0\rangle {\rm sgn(y_0)}.
\end{equation}
The right-hand form shows that this information on causality is contained in the 
propagator.

We will further 
consider that 
\begin{equation}
y_0 < z,
\end{equation}
so that if superluminality were concentrated on the AdS boundary, there is 
simply no time to causally propagate through the bulk to get to the 
boundary to take advantage of it. 
Therefore we are guaranteed 
that if  $\Delta H$ gives a non-zero 
correction to (\ref{order}),
 then {\it it must be due to superluminal effects in the 
gravitating AdS bulk spacetime}. 

Let  us check that there is indeed a non-vanishing correction to 
the bulk commutator VEV. First let us formalize the question in the 
gravity description. To do this let us adopt a Hamiltonian approach to the
AdS gravitational theory, using the AdS asymptotics to define energy and 
hence the Hamiltonian, $H_{AdS}$. The AdS/CFT dictionary tells us how 
insertions of the local operator ${\cal O}$ map to the AdS boundary,
so that we are able to map the effects of our CFT deformation, $\Delta H$, 
and regard it as a deformation added to $H_{AdS}$ (at least to any fixed order
in $\epsilon$ perturbation theory). Below we will work in the gravity 
description, and the appearance of ${\cal O}$ or 
$\Delta H$ will always denote the corresponding objects mapped to the 
gravity side. So for example, the total gravity-side Hamiltonian is
\begin{equation} 
H_{gravity} = H_{AdS} + \Delta H.
\end{equation}
In this canonical approach let $\phi(\vec{x},z)$ 
denote the Schrodinger operator dual to ${\cal O}$, with 
$\hat{\phi}(x,z) \equiv e^{iH_{gravity}x_0} \phi(\vec{x},z) 
e^{-iH_{gravity}x_0}$ being the usual Heisenberg operator construction. 
Using  the very general result of (\ref{pert}) it 
is then straightforward to prove that to first order in $\Delta H$,
\begin{eqnarray}
[\hat{\phi}(y,z), \hat{\phi}(0,z)] &=& i \int_0^{y_0} d \tau 
[[\hat{\phi}(y,z), \Delta \hat{H}(\tau)], \hat{\phi}(0,z)],
\end{eqnarray}
where we have used that at zeroth-order (relativistic) causality implies
 $[\hat{\phi}(y,z), \hat{\phi}(0,z)] = 0$. 

We can then calculate the VEV of this commutator in the large-$N_{color}$ 
limit, where the resulting four-point VEVs factorize as 
$\langle 0|\hat{\phi} \hat{\phi} \hat{\cal O} \hat{\cal O} |0 \rangle \approx 
\langle 0|\hat{\phi} \hat{\cal O} |0 \rangle 
\langle 0|\hat{\phi} \hat{\cal O} |0 \rangle$, etc. The result is 
\begin{eqnarray}
&~& \langle 0| [\hat{\phi}(y,z), \hat{\phi}(0,z)] | 0\rangle \nonumber \\
&\approx&
2i \int_0^{y_0} d \tau \int d^4 x d^4 x' J(x) J(x') 
\langle 0| [\hat{\cal O}(x_0 + \tau, \vec{x}), \hat{\phi}(y,z)] |0 \rangle 
\langle 0| [\hat{\cal O}(x'_0 + \tau, \vec{x}'), \hat{\phi}(0,z)] |0 \rangle
 \nonumber \\
&=& - 8i \int_0^{y_0} d \tau \int d^4 x d^4 x' J(x) J(x') 
{\rm Im} 
\langle 0| T \{ \hat{\cal O}(x_0 + \tau, \vec{x}) \hat{\phi}(y,z)] \} 
|0 \rangle  {\rm sgn}(x_0 + \tau -y_0) \nonumber \\
&~& \times
{\rm Im} \langle 0| T \{ [\hat{\cal O}(x'_0 + \tau, \vec{x}') 
\hat{\phi}(0,z)] \} |0 \rangle {\rm sgn}(x'_0 + \tau) \nonumber \\
&\equiv& -8i \int_0^{y_0} d \tau \int d^4 x d^4 x' J(x) J(x') 
{\rm Im} K(y_0- x_0 - \tau, \vec{y} -\vec{x}, z) 
{\rm Im} K(- x'_0 - \tau, -\vec{x}', z) \nonumber \\
&~& \times {\rm sgn}(x_0 + \tau -y_0) ~{\rm sgn}(x'_0 + \tau),
\end{eqnarray}
finally arriving at an expression in terms of
bulk-boundary propagators of the undeformed theory. 

Choose some intermediate $\tau \approx y_0/2$ as an example.
We see that if the support of $J$ has sufficiently positive $x'_0$ so that 
$(x'_0 + \tau)^2 > (\vec{x}')^2 + z^2$ and has sufficiently negative $x_0$ 
so that $(y_0 - \tau -x_0)^2 > (\vec{y} - \vec{x})^2 + z^2$, then 
causal communication between the bulk point $(0,z)$ and boundary point 
$(x'_0 + \tau, \vec{x}')$ is possible and causal 
communication between the bulk point $(y,z)$ and boundary point 
$(x_0 + \tau, \vec{x})$ is also possible. Therefore by the 
relation  between commutator VEVs and propagators (the analog of 
(\ref{order})) each Im$K$ factor can be non-zero and so we have demonstrated
 causal 
communication between $(0,z)$ and $(y,z)$.

It is important to stress that although we have expressed this 
communication mathematically as a  product of communication from 
bulk to boundary and then back in the undeformed theory, there is in fact not 
enough time available in our set-up 
for bulk to boundary communication to proceed unless there is superluminality 
in the bulk. This bulk superluminal communication must therefore 
be taking place. 
It is a mere convenience that we are parametrizing the requisite 
bulk disturbances in terms of 
 boundary sources that could produce them given enough time.

In this, admittedly indirect, manner we have shown that superluminality is 
taking place in the AdS bulk, and therefore the exotic Higgs effect necessary 
to make this possible in a gravitating spacetime must be present. 
But we must still ask if the gravitating theory is 
UV-complete. This is guaranteed if the deformed CFT is UV complete. 
We now show this.

\subsection{UV finiteness}

Consider the
amplitude to evolve from a state $|A\rangle$ to a state $|B\rangle$ over a 
time interval $t$. 
We consider the two states to be energy-momentum
 eigenstates of the undeformed CFT.
For clarity, we begin with the example of the order $\epsilon^2$ 
contribution to this amplitude following from (\ref{pert}), 
\begin{eqnarray}
&\propto&  \int_0^t d \tau' \int^t_{\tau'} d \tau  
\int d^4 x  d^4 y  d^4 x' d^4 y' J(x) J(y) J(x') J(y') \nonumber \\
&~& \times  \langle B| \hat{\cal O}(x_{0} + \tau, \vec{x}) 
\hat{\cal O}(y_{0} + \tau, \vec{y}) 
 \hat{\cal O}(x'_{0} + \tau', \vec{x}') 
\hat{\cal O}(y'_{0} + \tau', \vec{y}') |A\rangle 
\nonumber \\
&\propto&  \int d \tau' \int_{\tau'} d \tau 
\int d^4 x  d^4 y  d^4 x' d^4 y' J(x) J(y) J(x') J(y') 
\int_+ d^4 p  \int_+ d^4 q \int_+ d^4 k\nonumber \\
&~& \times  \langle B| \hat{\cal O}(x_{0} + \tau, \vec{x}) |p \rangle \langle p| 
\hat{\cal O}(y_{0} + \tau, \vec{y}) |q\rangle \langle q| 
 \hat{\cal O}(x'_{0} + \tau', \vec{x}') |k \rangle \langle k| 
\hat{\cal O}(y'_{0} + \tau', \vec{y}') |A\rangle 
\nonumber \\
&=& \int d \tau' \int_{\tau'} d \tau  
 \int_+ d^4 p  \int_+ d^4 q 
\int_+ d^4 k \tilde{J}(p_B-p)  \tilde{J}(p-q) \tilde{J}(q-k) \tilde{J}(k-p_A) 
\nonumber \\
&~& \times
e^{i (p_{B0}- q_0) \tau} e^{i (q_0 -p_{A0}) \tau'} 
\langle B| \hat{\cal O}(0) |p \rangle \langle p| \hat{\cal O}(0) | q \rangle 
\langle q | 
\hat{\cal O}(0) | k \rangle \langle k | \hat{\cal O}(0) | A \rangle.
\end{eqnarray}
We have inserted complete sets of states between 
operators, explicitly summing over their possible momenta, with positive 
mass-squared and energy (the $''+''$ subscript on the momentum integrals), 
and implicitly over any other labels. Such integrals represent sums over 
non-vacuum states related by Poincare symmetry and scale symmetry. One can 
also insert the vacuum state, in which case the relevant momentum integral 
drops out in the obvious way. We have not written these terms because they are 
less dangerous to finiteness.

UV divergences can only arise in the expression 
 when the momentum integrals or $\tau$ integrals diverge.
There are no UV divergences in the above expression however, 
because the smoothness 
of $J$ translates into the rapid damping of $\tilde{J}$ for large momenta, 
and because the $\tau, \tau'$ integrals have finite range  with only 
well-behaved phase factor integrands. This generalizes the finiteness we saw
in the second example of the last subsection.
The reader can easily extend this 
check of finiteness to arbitrary order in $\epsilon$ perturbation theory by 
repeated insertion of a complete set of states between operators. 
The smooth $J$ factors always make each momentum integral converge.

Thus our deformed CFT, and its deformed AdS dual, are UV complete.
We can easily compare this with Section 4 by noting that 
we revert to that case in the limit
\begin{eqnarray}
J(x) &\rightarrow& J(\vec{x}) \delta(x_0) \nonumber \\
\tilde{J}(q) &\rightarrow& \tilde{J}(\vec{q}).
\end{eqnarray}
Consider the simple case where the $|p\rangle, |k\rangle$ states are replaced 
by the CFT
 vacuum state, so there is only one momentum integral to worry about, 
and  only one (relative) $\tau_- \equiv \tau- \tau' > 0$ integral.
The relevant term is 
\begin{equation}
\int_0 
d \tau_- \int_+ d^4 q e^{i q_0 \tau_-} |\langle 0| {\cal O} |q \rangle
\tilde{J}(\vec{q})|^2.
\end{equation}
By scale invariance, $|\langle 0| {\cal O} |q \rangle|^2 \propto (q^2)^{d -2}$,
and we see that without $\tilde{J}$ to help cut off $q_0$, the integral
diverges as $\tau_- \rightarrow 0, q_0 \rightarrow \infty$.
 
\subsection{Perturbativity and vacuum decay}

Schematically, each order in perturbation theory in $\epsilon$ brings an 
expression,
\begin{equation}
\epsilon \int d \tau \int d^4 x J(x) \int d^4 y J(y) \int_+ d^4 p \int_+ d^4 q
 ... \hat{\cal O}(x_0 + \tau, \vec{x}) |p\rangle \langle p| 
\hat{\cal O}(y_0 + \tau, \vec{y}) |q\rangle \langle q|...
\end{equation}
The coupling combination $\epsilon J(x) J(y)$ has dimension $9-2d$ which 
we will ascribe completely to $\epsilon$, taking $J$ to be a smooth 
dimensionless 
function taking values of order unity in a spacetime volume of order $L^4$.
We have seen in the last subsection how the intermediate state momenta are 
tied to the external momenta, which we characterize to be of order $E$, 
with the $J$ integrals providing momentum shifts of order $1/L$. 
 We begin by considering $E \gg 1/L$. Each $J$ integral suppresses 
one $\int_+ d^4 p$ integral, fixing $p_{\mu} \sim E$. Each $\hat{\cal O}$ 
then counts as $E^d$ in its matrix elements and each $|p\rangle \langle p|$ 
counts as $E^{-4}$, by dimensional analysis and the fact that the hard 
external $E$ scale is dominant. Finally, the $\tau$-dependence in the 
operators turns into a phase factor $\sim e^{i \Delta E \tau}$, where 
$\Delta E$ is an energy change allowed by the ``background'' $J$, of order 
$1/L$. Therefore the $\int d \tau$ integral (with whatever time-ordered 
limits of integration) is at most of order $L$. 

Putting these factors together, we find that every order in perturbation theory
counts as the dimensionless combination, $\epsilon L E^{2d-8}$, 
for $E \gg 1/L$.
If $d > 4$ then, no matter how small we take $\epsilon L$, there 
will be processes where perturbation theory is breaking down (although 
we still do not find UV divergences perturbatively). We will avoid this 
by restricting the primary operator ${\cal O}$ to have scaling dimension
$d \leq 4$. Then the condition for perturbativity is 
$\epsilon L^{9-2d} \ll 1$ for $E > 1/L$. 

We must still check perturbativity in the far IR, $E \ll 1/L$. Now all 
momentum scales above are dominated by $1/L$ in all the integrals and matrix 
elements, and the 
perturbative strength is simply $\epsilon L^{9-2d}$. So there is consistency 
in the perturbativity requirements for IR and UV processes, $\epsilon L^{9-2d}
\ll 1$. 

Our deformation of SYM is finite and perturbative. In particular, the deformed 
Hamiltonian requires no subtractions and is a sum of squares ($H_{CFT}$ is 
by supersymmetry), so that there is some well-defined ground state. 
The  SYM vacuum is a finite and well-defined excitation above this 
ground state. We can make it arbitrarily long-lived 
by making $\epsilon$ weaker and weaker. Or we can do what we did in 
subsection 4.3, make $\epsilon \rightarrow \epsilon(\tau)$ and our 
Hamiltonian time-dependent so that the deformation is turned on for a 
finite duration. We then take 
$\epsilon$  small enough that the SYM (AdS) vacuum in 
most regions of space survives the period of deformation. This 
suppresses the amplitude for superluminality in the bulk but it can still 
take place.

\section{Conclusions}

Our central construction has been a weak Lorentz-violating deformation of 
 the ${\cal  N} = 4$ SYM CFT Hamiltonian by 
a superposition of spatially non-local operators. We 
checked that, 
at leading order in the perturbation, the standard AdS/CFT map 
gives a non-vanishing propagator between two spacetime points which are 
ordinarily causally disconnected in the AdS bulk. The two bulk points are also 
ordinarily out of causal contact with the AdS boundary during the time interval
separating them, implying that the 
superluminal behavior is taking place in the gravitating bulk.
Finally, we checked that our deformation was UV complete, in that 
there were no new sources of UV divergence outside the renormalized CFT. 
Therefore, there must be a complete deformed AdS gravity/string dual of 
the superluminal behavior. 

However, given our indirect CFT
approach to this conclusion, the detailed AdS description of bulk 
superluminality is not apparent. Indeed, the specific form of our spatially 
non-local deformation was chosen to demonstrate 
 bulk superluminality
 in terms of the simplest object of the AdS/CFT dictionary, namely
the bulk-boundary propagator. It is possible that a different 
non-local deformation might yield a simpler AdS spacetime description, 
although likely at the cost of a more complex translation from the CFT side.

We have presented a simple building block for superluminality, but there are
clearly other directions to pursue. It appears straghtforward that a similar 
deformation could be used to couple two different, otherwise decoupled CFTs, 
whose dual would describe {\it bulk} 
coupling of AdS degrees of freedom from both CFTs. 
Deformations by local operators (as opposed to non-local ones such as 
 we are suggesting) connecting otherwise disconnected CFTs have been discussed
in Refs. \cite{product}.
More ambitiously, 
we would like to understand regimes in which the Lorentz-violating or 
superluminal effects become important, 
as for example required in Ref. \cite{us}, 
in resolving the cosmological constant problem by Energy-Parity, or in 
\cite{bh}  in modifying the character of black hole horizons.
 We would 
also like to see if superluminal effects can be engineered as weak probes of 
ordinary horizons, as well as the interesting singularities that they can hide.

As discussed in the introduction, Lorentz violation in a gravitational 
context must appear formally as a type of Higgs effect. This suggests
 that even when the 
violation is explicit on the CFT side as in our case, on the AdS side it should
appear as a property of a {\it state} or solution, not a modification of the 
gravitational dynamics itself. As remarked above, our approach does not 
straightforwardly give a detailed description of such states in AdS.
It is intriguing however that, from the opposite 
direction, wormhole 
solutions in Euclidean AdS gravity pose a puzzle for CFT interpretation 
\cite{rey} \cite{juan} \cite{nima}
precisely because they suggest non-local interactions on the CFT side. Perhaps 
 the resolution of this puzzle lies in non-local deformations of the CFT, 
at least similar in spirit to the example of Section 5 of this paper.

The example of superluminality and Lorentz-violation 
provided in this paper has a certain 
``premeditated'' feel to it, and one naturally wonders whether it is too 
contrived to be at work in Nature. That would however be a premature 
conclusion, because we have only given an existence proof. It is possible that
real world gravity and relativity is a rich emergent phenomenon with a more 
 natural framework for these exotic effects. Hopefully we can 
understand the theoretical possibilities well enough to devise the right 
experimental tests to decide.

\section*{Acknowledgements}

The author is grateful to Nima Arkani-Hamed, Juan Maldacena and Joe Polchinski 
for discussions.
This research was supported by the National Science Foundation grant 
NSF-PHY-0401513 and by the Johns Hopkins Theoretical Interdisciplinary 
Physics and Astrophysics Center.

\end{document}